\documentclass[aps,prl,twocolumn,superscriptaddress,showpacs,preprintnumbers]{revtex4}

\usepackage{graphicx}

\begin{document}
\title{Identification of Nodal Kink in Electron-Doped (Nd$_{1.85}$Ce$_{0.15}$)CuO$_4$ Superconductor
from Laser-Based Angle-Resolved Photoemission Spectroscopy}

\author{Haiyun Liu$^{1}$, Guodong Liu$^{1}$, Wentao Zhang$^{1}$,
Lin Zhao$^{1}$, Jianqiao Meng$^{1}$, Xiaowen Jia$^{1}$,  Xiaoli
Dong$^{1}$, Wei Lu$^{1}$, Guiling Wang$^{2}$, Yong Zhou$^{2}$, Yong
Zhu$^{3}$, Xiaoyang Wang$^{3}$, Tao Wu$^{4}$, Xianhui Chen$^{4}$, T.
Sasagawa$^{5}$, Zuyan Xu$^{2}$, Chuangtian Chen$^{3}$ and X. J.
Zhou$^{1,*}$ }

\affiliation{
\\$^{1}$National Laboratory for Superconductivity, Beijing National Laboratory for
Condensed Matter Physics, Institute of Physics, Chinese Academy of
Sciences, Beijing 100190, China
\\$^{2}$Laboratory for Optics, Beijing National Laboratory for Condensed Matter Physics,Institute
of Physics, Chinese Academy of Sciences, Beijing 100190,  China
\\$^{3}$Technical Institute of Physics and Chemistry,
Chinese Academy of Sciences, Beijing 100190, China
\\$^{4}$Department of Physics, University of Science and Technology of China, Anhui, China
\\$^{5}$Materials and Structures Laboratory, Tokyo Institute of Technology, Yokohama Kanagawa, Japan
 }

\date{August 5, 2008}

\begin{abstract}

High-resolution laser-based angle-resolved photoemission
measurements have been carried out on the electron-doped
(Nd$_{1.85}$Ce$_{0.15}$)CuO$_4$ high temperature superconductor. We
have revealed a clear kink at $\sim$60 meV in the dispersion along
the (0,0)-($\pi$,$\pi$) nodal direction, accompanied by a
peak-dip-hump feature in the photoemission spectra. This indicates
that the nodal electrons are coupled to collective excitations
(bosons) in electron-doped superconductors, with the phonons as the
most likely candidate of the boson. This finding has established a
universality of nodal electron coupling in both hole- and
electron-doped high temperature cuprate superconductors.

\end{abstract}


\maketitle

High temperature superconductivity in cuprates is achieved by doping
an appropriate amount of charge carriers (electrons or holes) into
the parent antiferromagnetic Mott insulators\cite{LeeReview}. It is
known that the unusual physical properties of high-T$_c$ cuprate
superconductors stem from the complex interplay between electron
charge, spin and lattice vibrations. Investigation of these
many-body effects is essential to understand the macroscopic
physical properties and the mechanism of high temperature
superconductivity. High resolution angle-resolved photoemission
spectroscopy (ARPES) has become a powerful tool to probe many-body
effects because the interaction of electrons with a collective
excitation gives rise to a change in the electron self-energy which
can be measured directly from ARPES\cite{ThreeReviews}. In
hole-doped cuprates, ARPES measurements have revealed a ubiquitous
existence of a kink in the dispersion at $\sim$70meV along the
(0,0)-($\pi$,$\pi$) nodal direction signaling an electron coupling
with low energy collective excitations (phonons or magnetic
resonance
mode)\cite{Bogdanov,PJohnson,Kaminski,Lanzara,XJZhou,Kordyuk}.
However, in the electron-doped cuprates,  the ARPES measurements so
far have not found indication of such a kink in the nodal
dispersion\cite{TSato,ArmitageKink,Matsui013}.  It is well-known
that the electron-doped superconductors exhibit different behaviors
from its hole-doped counterparts, such as the narrower
superconducting doping range and lower superconducting transition
temperature (T$_c$)\cite{Tokura}. This raises interesting questions
on whether the nodal electron dynamics in the electron-doped
cuprates is inherently different from the hole-doped ones and
whether such a difference may be responsible for the distinct
behaviors between the electron- and hole-doped
cuprates\cite{ZXShen}.

In this paper we report an identification of a kink in the nodal
dispersion of the electron-doped (Nd$_{1.85}$Ce$_{0.15}$)CuO$_4$
superconductor by taking advantage of the high performance
laser-based ARPES system. In both optimally-doped and underdoped
(Nd$_{1.85}$Ce$_{0.15}$)CuO$_4$ superconductors, we have clearly
identified a kink in the nodal dispersion near 60 meV, accompanied
by peak-dip-hump structure in the photoemission spectra and a drop
in the quasiparticle scattering rate. These observations indicate
that, in the electron-doped cuprates, the nodal electrons couple
with some collective excitations with phonons as the most likely
candidate.  It has established a universality of the nodal kink in
both electron- and hole-doped cuprate superconductors. It also
suggests that the different behaviors between electron- and
hole-doped cuprates may not be dictated by the nodal electron
coupling, but by their distinct doping-dependent electronic
structure.

The angle-resolved photoemission measurements were performed on our
recently developed vacuum ultra-violet (VUV) laser-based ARPES
system at the Institute of Physics, Chinese Academy of Sciences,
which has some unique advantages such as super-high energy
resolution (better than 1 meV), high momentum resolution, super-high
photon flux and enhanced bulk sensitivity\cite{LiuLaser}. The photon
energy of the VUV laser is 6.994 eV with a bandwidth of 0.26 meV.
Because of the relatively weaker photoemission cross section of the
electron-doped (Nd$_{1.85}$Ce$_{0.15}$)CuO$_4$ samples, the energy
resolution of the electron energy analyzer (Scienta R4000) was set
at 6.25 meV, giving rise to an overall energy resolution of 6.26
meV.  The angular resolution is $\sim$0.3$^\circ$, corresponding to
a momentum resolution $\sim$0.004 $\AA$$^{-1}$ at the photon energy
of 6.994 eV.  The Fermi level is determined by referencing to the
Fermi edge of a clean polycrystalline gold that is electrically
connected to the sample. The
(Nd$_{1.85}$Ce$_{0.15}$)CuO$_{4+\delta}$  single crystals were grown
by flux method\cite{XHChen} or floating zone furnace. By annealing
the as-grown single crystals in argon atmosphere to control the
oxygen content, $\delta$, we obtained two kinds of superconducting
samples, one is close to optimally-doped with a T$_c$=24 K while the
other is slightly underdoped with a T$_c$=20 K. The single crystals
were cleaved {\it in situ} and measured in ultra-high vacuum with a
base pressure better than 5$\times$10$^{-11}$ Torr.

\begin{figure}[tbp]
\begin{center}
\includegraphics[width=1.0\columnwidth,angle=0]{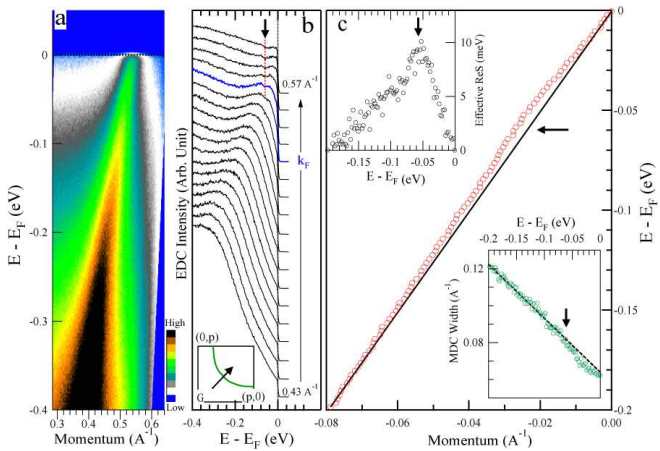}
\end{center}
\caption{Nodal kink in the optimally-doped
(Nd$_{0.85}$Ce$_{0.15}$)CuO4 (T$_c$=24 K) sample. (a). Photoemission
image along the (0,0)-($\pi$,$\pi$) nodal direction at 13 K. The
location of the momentum cut is shown in the inset of (b). (b). The
corresponding photoemission spectra. The EDCs near the Fermi
momentum shows a peak-dip-hump structure, with the dip position
marked by a dashed red line. (c). Nodal dispersion extracted from
the MDC analysis. The black straight line is a guide to the eye
which also serves as an empirical bare band to extract the effective
real part of electron self-energy (Re$\Sigma$), as shown in the
upper-left inset.  The bottom-right inset shows the MDC width. The
black dashed line is a guide to the eye. }
\end{figure}

Fig. 1 shows the photoemission data of the optimally-doped
(Nd$_{0.85}$Ce$_{0.15}$)CuO4 (T$_c$=24 K) sample measured along the
(0,0)-($\pi$,$\pi$) nodal direction at a temperature of 13 K. A
clear band dispersion is observed in the raw data (Fig. 1a). By
fitting the corresponding MDCs (momentum distribution curves),  the
quantitative dispersion is obtained as plotted in Fig. 1c. A weak
but clear kink is present in the dispersion at an energy of $\sim$60
meV, as indicated by the arrow in Fig. 1c. By choosing the straight
line connecting the two points in the dispersion at E$_F$ and -0.2
eV as an empirical bare band, the corresponding ``effective real
part of electron self-energy (Re$\Sigma$)" is extracted (up-left
inset of Fig. 1c) with the $\sim$60 meV feature showing up more
clearly. In the corresponding MDC width (bottom-right inset of Fig.
1c), which is related to the quasiparticle scattering rate or the
imaginary part of the electron self-energy, a slight drop is
discernable near $\sim$60 meV, as seen in the deviation of the MDC
width data from a linear line. In the corresponding photoemission
spectra (energy distribution curves, EDCs) as shown in Fig. 1b, the
EDCs near the Fermi momentum show clear double-peak or peak-dip-hump
structure, with a location of the dip at $\sim$60 meV, as indicated
by the dashed red line in Fig. 1b.

The photoemission data for the slightly underdoped
(Nd$_{0.85}$Ce$_{0.15}$)CuO4 (T$_c$=20 K) sample (Fig. 2) show some
similar features as those in the optimally-doped sample (Fig. 1).
First, a clear kink near $\sim$60 meV is present in the MDC-derived
dispersion (Fig. 2c) and more clearly in the corresponding effective
real part of the electron self-energy (up-left inset of Fig. 2c).
Second, in the corresponding EDCs near the Fermi momentum (Fig. 2b),
the peak-dip-hump structure can be seen with a dip energy at
$\sim$60 meV. Here because of the data scattering it is hard to
discern a clear feature in the MDC width (bottom-right inset of Fig.
2c) . An obvious difference in the data for the slightly underdoped
sample is the observation of another kink at lower binding energy,
$\sim$23 meV, as seen in the dispersion (Fig. 2c) and the
corresponding effective real part of electron self-energy (up-left
inset of Fig. 2c). This $\sim$23 meV kink arises because of the
existence of a gap along the nodal direction. This gap opening is
evidenced by the corresponding peak position of the EDC at the Fermi
momentum which is about 35 meV  below the Fermi level (the
corresponding EDC leading edge is about 12 meV below the Fermi
level) (Fig. 2b). The appearance of the low energy kink and nearly
vertical dispersion is similar to that found in the hole-doped
superconductors when a superconducting gap opens below
T$_c$\cite{NormanMDCGap}. Note that this nodal gap in the underdoped
(Nd$_{0.85}$Ce$_{0.15}$)CuO4 (T$_c$=20 K) sample is a band gap, but
not the superconducting gap which is zero along the nodal
direction\cite{TSato,ArmitageSCGap,MatsuiSCGap}.

\begin{figure}[tbp]
\begin{center}
\includegraphics[width=1.0\columnwidth,angle=0]{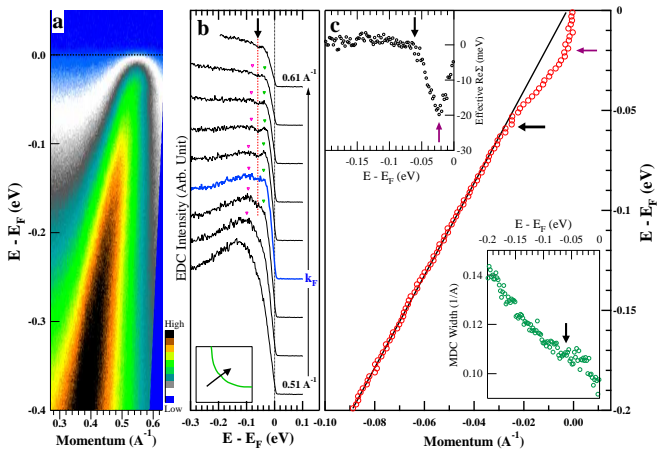}
\end{center}
\caption{Nodal kink in the slightly underdoped
(Nd$_{0.85}$Ce$_{0.15}$)CuO4 (T$_c$=20 K) sample. (a). Photoemission
image along the (0,0)-($\pi$,$\pi$) nodal direction at 15 K. The
location of the momentum cut is shown in the inset of (b). (b). The
corresponding photoemission spectra. The EDCs near the Fermi
momentum shows a peak-dip-hump structure, with the peak and hump
positions marked by triangles and the dip position marked by a
dashed red line. (c). Nodal dispersion extracted from the MDC
analysis. The black straight line is a guide to the eye which also
serves as an empirical bare band to extract the effective real part
of electron self-energy (Re$\Sigma$), as shown in the upper-left
inset. A kink in the dispersion as well as an abrupt change in the
effective Re$\Sigma$ can be identified at $\sim$60 meV, as indicated
by black arrows. Another kink near 23 meV can also be seen in the
dispersion and the effective Re$\Sigma$, as indicated by the purple
arrows. The bottom-right inset shows the MDC width.  }
\end{figure}

Fig. 3 shows momentum dependence of the $\sim$60 meV kink feature in
dispersions for the (Nd$_{0.85}$Ce$_{0.15}$)CuO4 samples. For the
optimally-doped sample (T$_c$=24 K) (Fig. 3a), the $\sim$60 meV kink
persists over a relatively large momentum space that is covered,
with the kink feature getting slightly weaker when the momentum move
away from the nodal region (Fig. 3a). Also note in this momentum
space, there is no indication of band gap opening, as seen from the
EDCs on the Fermi surface(Fig. 3b). For the slightly underdoped
sample (T$_c$=20 K) (Fig. 3c),  a band gap is present near the nodal
region which gets larger when the momentum moves away from the nodal
region, as seen from the position of EDCs on the Fermi surface (Fig.
3d). This observation is consistent with the previous
measurements\cite{NPAdoping,MatsuiDoping}. The gap size increase
shifts the energy position upwards for the kink induced by the band
gap (Fig. 3c). As shown in Fig. 3c, for the momentum cuts closest to
the nodal direction (Cuts 1 and 2), there is a clear $\sim$60 meV
kink in dispersions with the band-gap-induced kink confined to a low
energy ($\sim$23 meV). For the Cut 3, even though the
band-gap-induced kink energy gets higher ($\sim$34 meV) one can
still observe the $\sim$60 meV kink which appears to get weaker than
that for Cuts 1 and 2. For the Cut 4, the band-gap-induced kink has
a comparable energy scale at $\sim$60 meV, making the $\sim$60 meV
kink hard to observe.

Fig. 4 shows the temperature dependence of the $\sim$60 meV nodal
kink in the optimally-doped (Nd$_{0.85}$Ce$_{0.15}$)CuO4 (T$_c$=24
K) sample. It is clear that the $\sim$60 meV kink is present over
the entire temperature range measured: both below T$_c$ and above
T$_c$. Particularly, there is little change observed across the
superconducting transition temperature.

\begin{figure}[bp]
\begin{center}
\includegraphics[width=1\columnwidth,angle=0]{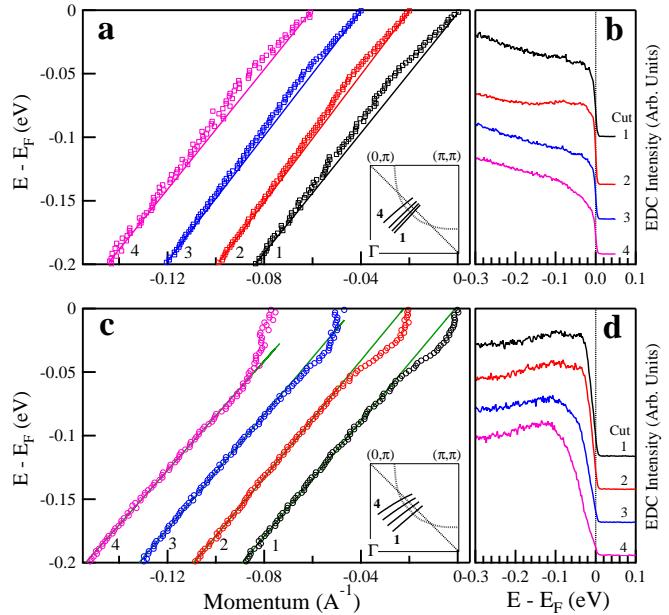}
\end{center}
\caption{Momentum dependence of the dispersion and photoemission
spectra on the Fermi surface in (Nd$_{1.85}$Ce$_{0.15}$)CuO$_4$
superconductors. (a). Momentum-dependent dispersions for the
optimally-doped (Nd$_{1.85}$Ce$_{0.15}$)CuO$_4$ (T$_c$=24 K) sample.
The solid lines are guides to the eye. The location of the momentum
cuts are marked in the bottom-right inset. For clarity, dispersions
are offset along the momentum axis. (b). EDCs on the Fermi surface
for the (Nd$_{1.85}$Ce$_{0.15}$)CuO$_4$ (T$_c$=24 K) sample. (c).
Momentum-dependent dispersions for the slightly underdoped
(Nd$_{1.85}$Ce$_{0.15}$)CuO$_4$ (T$_c$=20 K) sample. The solid lines
are guides to the eye. The location of the momentum cuts are marked
in the bottom-right inset.  For clarity, dispersions are offset
along the momentum axis.  (d). EDCs on the Fermi surface for the
(Nd$_{1.85}$Ce$_{0.15}$)CuO$_4$ (T$_c$=20 K) sample.
 }
\end{figure}

The above systematic measurements on the doping-, momentum- and
temperature-dependence of the nodal $\sim$60 meV kink provide
important information to judge on its origin in the electron-doped
(Nd$_{1.85}$Ce$_{0.15}$)CuO$_4$ superconductors. We note that,
although the nodal kink was not observed before in the
electron-doped superconductors, there have been reports on the
signatures of a kink near the ($\pi$, 0) antinodal
region\cite{TSato,ArmitageKink,Matsui013}. Such an antinodal kink is
attributed either to the band-folding effect\cite{Matsui013} or to
the mode coupling\cite{TSato,ArmitageKink}.  In the electron-doped
cuprates, the commensurate antiferromagnetic fluctuation is found to
be strong and may coexist with
superconductivity\cite{YamadaNeutron}, which may give rise to a band
folding with respect to the antiferromagnetic zone boundary. In this
case, the original band and the folded band cross, forming a gap at
the intersection, and separate into two (upper and lower) band
sections\cite{Matsui013}. Such a picture was proposed to account for
the kink and the peak-dip-hump structure near the antinodal
region\cite{Matsui013}. However, this scenario is not applicable to
the kink we have observed near the nodal region. In terms of the
band-folding picture, because the upper section of the band is well
above the Fermi energy along the nodal direction, the band folding
effect on the occupied state is expected to be
minimal\cite{Matsui013}. Particularly, it would not produce the
peak-dip-hump structure as we have observed in EDCs near the nodal
region (Figs. 1b and 2b).

The above observations of the nodal $\sim$60 meV kink in the
dispersion, a drop in the quasiparticle scattering rate and the
peak-dip-hump structure in EDCs in the electron-doped
superconductors show a clear resemblance to those found in the
hole-doped
counterparts\cite{Bogdanov,PJohnson,Kaminski,Lanzara,XJZhou,Kordyuk}.
These strongly suggest that the nodal $\sim$60 meV kink in the
electron-doped cuprates is due to the electron coupling with some
collective excitations (bosons). Concerning the nature of the
involved boson(s), the available candidates that are present in the
electron-doped cuprates are either magnetic resonance
mode\cite{NCCOReso} or phonons.   The possibility of the magnetic
resonance mode can be easily ruled out because its energy scale
($\sim$10 meV)\cite{NCCOReso} is far off from 60 meV.  This has left
phonons as the most viable candidate with its comparable energy
scale. The temperature dependence measurement (Fig. 4) lends further
support to the phonon scenario. In the electron-doped cuprates, the
oxygen vibrations in the CuO$_2$ planes produce phonon modes in the
energy range of 50$\sim$80 meV\cite{KangPhonon,XRayPhonon,MBraden}.
Particularly, the mode with the bond-stretching character has an
energy near 60 meV and exhibits an anomalous softening indicative of
strong electron-phonon coupling\cite{MBraden}. This makes the
electron coupling with this particular phonon mode the most likely
case for the nodal $\sim$60 meV energy scale in the electron-doped
cuprates.

Our present work has clearly established the nodal kink as a
universal feature for both the electron-doped and hole-doped
cuprates. Although the mode coupling strength\cite{LambdaNote} of
the electron-doped superconductors ($\lambda$$\sim$0.20 for the
optimally-doped (Nd$_{1.85}$Ce$_{0.15}$)CuO$_4$) is weaker than that
in the hole-doped counterparts ($\lambda$$\sim$0.55 for the
optimally-doped (La$_{1.85}$Sr$_{0.15}$)CuO$_4$\cite{XJZhou}), such
a quantitative difference may not account for the dramatically
different behaviors between the electron- and hole-doped
cuprates\cite{Tokura}. One most notable distinction between the
electron- and hole-doped cuprates is the different doping evolution
of the electronic structure. In the hole-doped cuprates, upon doping
the antiferromagnetic parent compound, the low-lying states first
develop near the nodal region while the antinodal region is gapped
at low dopings\cite{YoshidaArc,ZhouArc,KShenArc}.  In the
electron-doped case, by contrast, the antinodal state first develops
upon doping\cite{ArmitageDoping} while the nodal region is gapped at
low dopings (Figs. 2 and 3d).  Such a dichotomy in the electronic
structure may play more important role in dictating the different
behaviors between the electron-doped and hole-doped cuprates.

\begin{figure}[tbp]
\begin{center}
\includegraphics[width=1.0\columnwidth,angle=0]{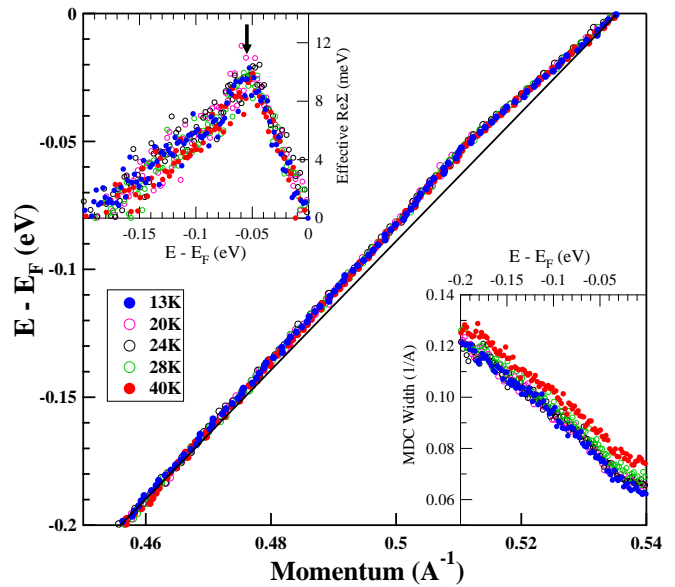}
\end{center}
\caption{Temperature dependence of the nodal dispersion in the
optimlly-doped (Nd$_{1.85}$Ce$_{0.15}$)CuO$_4$ (T$_c$=24 K) sample.
The black line is a guide to the eye which also serves as an
empirical bare band to extract the effective real part of electron
self-energy, Re$\Sigma$, as shown in the upper-left inset. The
bottom-right inset shows the MDC width measured at various
temperatures. }
\end{figure}

In summary, by taking high quality data from the laser-based ARPES,
we have clearly identified a  nodal kink at $\sim$60 meV in the
electron-doped cuprate superconductors.  This has established a
universality of the 60 meV nodal kink in both electron- and
hole-doped high temperature superconductors. The origin of the
$\sim$60 meV kink is attributed to the electron coupling with some
collective excitations, with the phonons being the most likely
candidate. This finding also suggests that the nodal electron
coupling may not be the major player for understanding different
behaviors between the electron- and hole-doped cuprates.

This work is supported by the NSFC, the MOST of China (973 project
No: 2006CB601002, 2006CB921302), and CAS (Projects ITSNEM).

$^{*}$Corresponding author: XJZhou@aphy.iphy.ac.cn

\newpage


\begin{thebibliography}{99}
\bibitem{LambdaNote} The empirical mode-coupling constant $\lambda$ is estimated by
$\lambda$=v$_{bare}$/v$_F$-1 where v$_{bare}$ represents the slope
of the straight line connecting the two points in the dispersion at
-0.2 eV and E$_F$ and v$_F$ the slope of the measured dispersion
between -0.05 eV and E$_F$.
\end{thebibliography}
\end{document}